\documentclass[aps,pre,twocolumn,groupedaddress,10 pt,showpacs,showkeys]{revtex4}
\usepackage{graphicx}
\usepackage{amsmath}
\usepackage{amssymb,color}
\usepackage[english]{babel}
\graphicspath{{c:/fortran/jose_maria/}}

\date{\today}

\begin{document}
\title{Extreme intensity pulses in a semiconductor laser with a short external cavity}
\author{Jose A. Reinoso}
\email{jmaparicio@bec.uned.es}
\affiliation{Departamento de Fisica Fundamental, Universidad Nacional de Educacion a Distancia (UNED), Paseo Senda del Rey 9, E-28040 Madrid, Spain}
\author{Jordi Zamora-Munt}
\email{jordi@ifisc.uib-csic.es}
\affiliation{IFISC (CSIC-UIB), Campus Universitat Illes Balears, E-07122 Palma de Mallorca, Spain}
\author{Cristina Masoller}
\email{cristina.masoller@upc.edu}
\affiliation{Departament de Fisica i Enginyeria Nuclear, Universitat Politecnica de Catalunya, Colom 11, ES-08222 Terrassa, Barcelona, Spain}

\begin{abstract}
We present a numerical study of the pulses displayed by a semiconductor laser with optical feedback in the short cavity regime, such that the external cavity round trip time is smaller than the laser relaxation oscillation period. For certain parameters there are occasional pulses, which are high enough to be considered extreme events. We characterize the bifurcation scenario that gives rise to such extreme pulses and study the influence of noise. We demonstrate intermittency when the extreme pulses appear and hysteresis when the attractor that sustains these pulses is destroyed. We also show that this scenario is robust under the inclusion of noise.

\end{abstract}
\date{\today}
\maketitle

\section{Introduction}

The study of extreme and rare events is a highly active and interdisciplinary research field \cite{prl_2005,pre_2012,pre_2011}. Extreme events can have catastrophic consequences in fields such as climatology, population dynamics or economy \cite{cc_2010,epj_2012,pre_2009}.
In lasers, for example, extreme and rare pulses have been observed in mode-locked lasers \cite{ol_2011} and in semiconductor lasers with continuous-wave optical injection \cite{prl_2011,el_2012} or with phase conjugated feedback \cite{ol_2013}.

We present a numerical study of the intensity pulses displayed by a semiconductor laser with optical feedback in the short cavity regime \cite{prl_2001,pre_2004,pra_2006}, such that the external cavity round trip time is smaller than the laser relaxation oscillation period. We use as a framework the well-known Lang-Kobayashi (LK) model~\cite{lk,libro_deb_alan,libro_othsubo}. Previous numerical work based on the LK model has found high intensity pulses in the laser chaotic output, that correspond to transitions between external cavity modes (ECMs)~\cite{pre_2004}. We characterize these pulses and show that in specific parameter regions they are high enough to be considered extreme events.

In extreme value analysis the definition of an extreme event is arbitrary as is associated to an event that is rare and that has an extreme deviation from the average. Qualitatively, extreme values are those in the tail of a long-tailed distribution; quantitatively, there are two main approaches to define extreme values: 1) values that exceed (or fall below) a certain threshold are considered extreme \cite{RogueBook}, and 2) maxima (or minima) in ``blocks'' of the time series are considered extreme \cite{ExtremeBook}. In this work we use the first criterion and define extreme intensity pulses as those above a certain threshold. One should notice that both approaches involve a certain degree of arbitrariness, either in the selection of the threshold or in the selection of the length of the block. An example of the use of the first criterium is in oceanography, where extreme waves (referred to as freak or rogue waves) are those whose height is larger than the mean value plus four to eight times the standard 
deviation of the height distribution, or as waves with abnormality index larger than 2 \cite{RogueBook}. An example of the use of the second criterium is in climate data analysis, where extreme values can be annual, biannual, etc.

Using the point-over-threshold criterium to define extreme intensity pulses, we study how they develop and how they are affected by noise. We demonstrate that an abrupt expansion in phase space of an attractor developed from an ECM creates an expanded attractor that sustains extreme pulses. For certain parameters this attractor coexists with a smaller attractor that develops from a different ECM. We identify two phenomena involved in the appearance and in the destruction of the attractor that sustains extreme pulses: deterministic intermittency when the attractor abruptly expands, and hysteresis when the attractor is destroyed. We also show that this scenario is robust under the inclusion of noise.

This paper is organized as follows. Section II briefly describes the model employed, which is the well-known delay-differential Lang-Kobayashi model~\cite{lk}. Section III presents the numerical results; we first focus on deterministic simulations and then discuss the influence of noise. Section IV presents a summary of the results and the conclusions.

\section{Model}

The rate equations for the complex optical field, $E$, and the excess carrier number, $N$, are~\cite{lk}
\begin{eqnarray}
 dE/ds &=& (1+i \alpha)N E(s)+\eta e^{-iw\theta} E(s-\theta) +\beta \xi \\
\label{eqn:E2}
 T dN/ds  &=& J-N-(1+2N)|E(s)|^{2}
 \label{eqn:N2}
\end{eqnarray}

In these equations the dimensionless time, $s$, and the delay time, $\theta$, are in units of photon lifetime $\tau_{p}$: $s=t/\tau_{p}$, $\theta=\tau/\tau_{p}$. The parameters are: $T=\tau_{n}/\tau_{p}$ where  $\tau_{n}$ is the carrier lifetime, the feedback rate is $\eta$, the feedback phase is $w\theta$, the pump current parameter is $J$ and the line width enhancement factor is $\alpha$. Spontaneous emission noise is taken into account by a complex additive Gaussian white noise, $\xi$, and the noise strength is $\beta$.

The time-delayed feedback renders the system multistable and the model has several fixed-point solutions, usually referred to as {\it external cavity modes} (ECMs), that can be calculated from
\begin{eqnarray}
\Delta \phi_{s}&=&-\eta\theta\sqrt{1+\alpha^{2}}\sin(\Delta \phi_{s}+w\theta+\arctan(\alpha))\\
\label{eqn:Aphi1}
E_{0s}^{2}&=&\frac{J-N_{s}}{1+2N_{s}}\\
\label{eqn:E0}
N_{s}&=&-\eta \cos(\Delta \phi_{s}+w\theta)
\label{eqn:N}
\end{eqnarray}

where $\Delta \phi_{s}$, $E_{0s}$ and $N_{s}$  are the steady state values of the phase difference, field amplitude and carrier number.
The number of ECMs increases with the feedback strength and their stability depends on the model parameters.

As $\eta$ increases new ECMs appear in pairs after saddle-node bifurcations. The initially stable ECMs leads a chaotic attractor after a series of bifurcations. With further increase of the feedback the chaotic attractors expand and eventually merge with previously existing attactors forming either an attractor ruin or a stable attractor \cite{sano,masoller,gavrielides}. In the first case the chaotic dynamics is a transient after which the trajectory finds a stable ECM; in the second case the attractor merging process results in a single stable attractor that can either coexist with a stable ECM, or it can be the only stable attractor (this occurs when there are no stable ECMs). When the chaotic dynamics is transient, it has been shown that the average duration of the transient increases several orders of magnitude for each unit we increase the value of the $\alpha$ parameter \cite{torcini,zamora}. Here we use the same parameters as in Ref.~\cite{pre_2004}, and the value of $\alpha$ is high
enough to guarantee that the duration of the chaotic transient (if any) is in practice infinite.

\section{Results}

We chose parameters similar to those in Ref.~\cite{pre_2004}, where the bifurcations of the ECMs were studied in detail: $T=1710$, $J=1.155$, $\alpha=5$ and $\theta=70$. In the simulations the feedback strength, the feedback phase and the noise strength are taken as control parameters. The equations are integrated with a second-order Runge-Kutta method with integration step $ds=0.01$. The initial conditions are such that the complex field and carrier number are close to zero. In the following we present first the results of deterministic simulations and then we discuss the influence of noise.

\subsection{Deterministic simulations}

\begin{figure}[]
\includegraphics[width=6 cm]{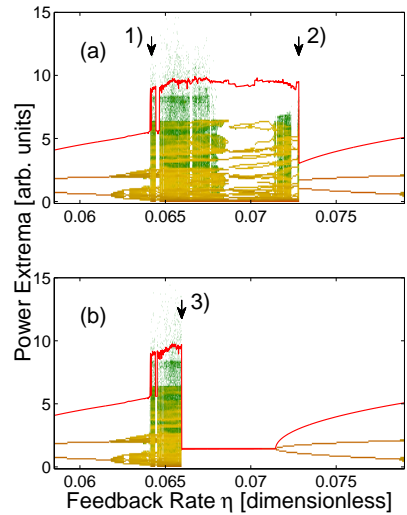}
\caption{(Color online) Bifurcation diagram when the feedback strength, $\eta$, increases (a) and then decreases (b). The pulse amplitude is plotted vs $\eta$, the horizontal solid line (red online) determines the threshold for a pulse to be considered an extreme event (equal to $5$ times the standard deviation over the mean value). The color gray code (brown-green online) in logarithmic scale indicates the number of pulses. Darker colors (brown online) indicate high-probable pulse amplitudes while lighter colors (green online), less probable amplitudes. The labels $1$, $2$ and $3$ indicate the transitions discussed in the text. $\beta=0$, other parameters as indicated in the text. \label{fig:1}}
\end{figure}

When the feedback strength varies the laser intensity displays a complicated sequence of bifurcations. In Fig.~\ref{fig:1} we plot the amplitude of the intensity pulses and the color code indicates in logarithmic scale the number of pulses, for increasing [Fig.~\ref{fig:1} (a)] and for decreasing [Fig.~\ref{fig:1} (b)] feedback. The bifurcation scenario is as discussed in~\cite{pre_2004}, and a similar one has been observed with opto-electronic feedback~\cite{epj_2010}. When the feedback increases, if $\eta < 0.064$ the intensity pulses are relatively small (below 4); however, slightly above this feedback level a sudden abrupt expansion of the pulse amplitude occurs reaching amplitudes higher than $10$, without a clear maximum. This abrupt change occurs at $\eta \sim 0.064$ and will be referred to as {\it transition 1}. By further increasing the feedback strength different dynamical regimes including periodic windows, chaos, and regular pulse packages (RPPs) occur~\cite{prl_2001,pre_2004,pra_2006}.

At {\it transition 1} the number of pulses (in color code) reveals that the pulse amplitudes are highly heterogeneous. There is a more densely visited range at low amplitudes that is similar before and after the expansion while large amplitude pulses occur only sporadically.

\begin{figure}[]
\includegraphics[width=6cm]{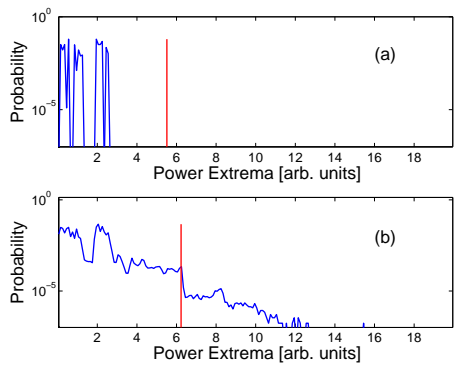}
\caption{(Color online) Distribution of pulse amplitudes for $\eta=0.064$ (a) and $\eta=0.064095$ (b), before and after respectively the attractor expansion ({\it transition 1}). The vertical line indicates the threshold over which extreme pulses are defined, which is equal to the mean value pulse height plus five times the standard deviation of the pulse height distribution.
\label{fig:pdf}}
\end{figure}

The distribution of pulse amplitudes, presented in Fig.~\ref{fig:pdf}, has a long-tail after the expansion [Fig.~\ref{fig:pdf}(b)] (which reveals the existence of extreme values), and a well-defined cut off before the expansion [Fig.~\ref{fig:pdf}(a)].

In order to characterize the extreme pulses, they will be defined quantitatively as those whose amplitude is larger than the mean value plus five times the standard deviation, $\sigma$, of the pulse amplitude distribution (the threshold is indicated with a red line in Figs.~\ref{fig:1} and ~\ref{fig:pdf}). As discussed in the Introduction, the criterion for defining quantitatively extreme pulses is quite arbitrary; however, the parameter region where extreme pulses are observed does not significantly change when the threshold is varied within the range of $5-8\sigma$. We use 5$\sigma$ as a compromise solution to have good statistics without having to perform extremely long simulations (to observe a significant number of extreme pulses defined with a higher threshold would require much longer simulations).

\begin{figure}[]
\includegraphics[width=6cm]{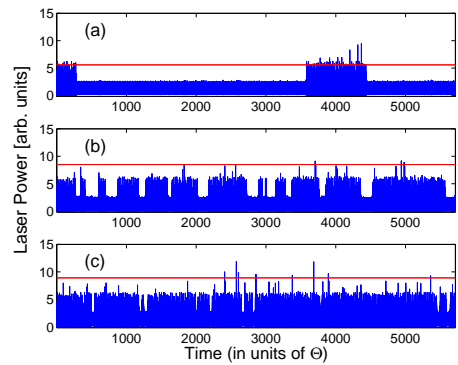}
\caption{Deterministic intermittency after {\it transition $1$} ($\beta=0$). $\eta=0.064095$ (a), $\eta=0.06412$ (b) and $\eta=0.0642$ (c).
\label{fig:3}}
\end{figure}

Immediately after the expansion of the attractor ({\it transition 1}) deterministic intermittency occurs as shown in Fig.~\ref{fig:3}, where alternating intervals of high and regular intensity pulses are seen; during the intervals where the laser displays high pulses, only a few of these pulses are extreme, i.e., cross the threshold represented by the horizontal line.

\begin{figure}[]
\includegraphics[width=6cm]{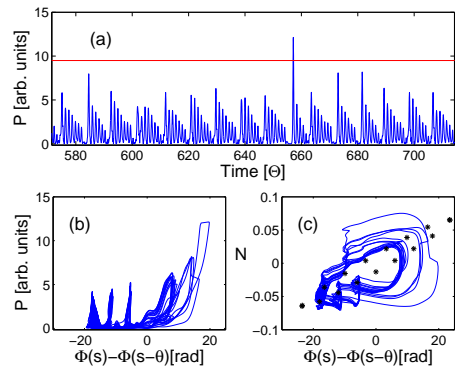}
\caption{Intensity time series where an extreme pulse is observed (a). Panels (b) and (c) display the phase portrait $[\Phi(s)-\Phi(s-\theta)],P]$, and $[\Phi(s)-\Phi(s-\theta),N]$. $\eta=0.066$, other parameters as in Fig. 1. \label{fig:2}}
\end{figure}

A detail of an extreme pulse is shown in Fig.~\ref{fig:2}(a), and the projections of the trajectory in the planes (Intensity, Phase delay) and (Carrier excess, Phase delay) are shown in Figs.~\ref{fig:2}(b) and Fig.~\ref{fig:2}(c) respectively. Figure~\ref{fig:2}(c) also displays the position of the ECMs. As can be seen in this figure, high pulses occur when the trajectory reaches large positive phase delays, $\Delta \phi_{s}$, while the regular-amplitude pulses do not approach that region of the phase space.

\begin{figure}[]
\includegraphics[width=6cm]{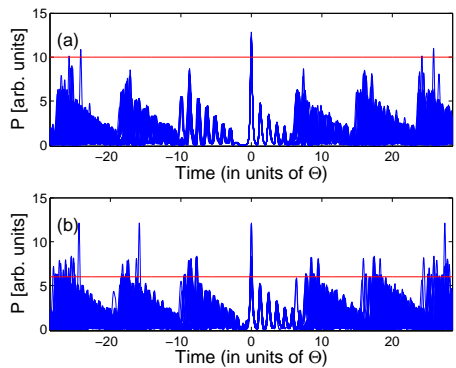}
\caption{(Color online) Superposition of extreme pulses, such that the time traces are centered at the peak of the pulse (see text for details). The feedback strength is $\eta = 0.066$, and other parameters are as indicated in the text. The threshold for defining extreme pulses, indicated with an horizontal line (red online), is equal to $10$ in (a) and is equal to $6$ in (b). In both panels the number of pulses is 52. \label{fig:4}}
\end{figure}

\begin{figure}[]
\includegraphics[width=6cm]{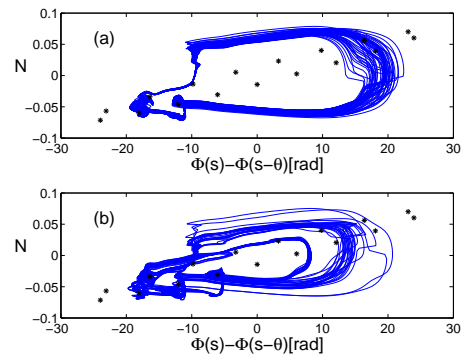}
\caption{Phase portrait $[\Phi(s)-\Phi(s-\theta),N]$ displaying the superposition of the sections of the trajectory that contain the 52 pulses shown in Fig.~\ref{fig:4}. In panel (a) the amplitude threshold is equal to 10, while in panel (b), is equal to 6.\label{fig:4b}}
\end{figure}

An abrupt expansion of pulse amplitudes similar to the one observed in {\it transition 1} was reported in a semiconductor laser with cw optical injection~\cite{pra_2013}. In~\cite{pra_2013} the attractor expansion was interpreted as due to the crossing of the attractor with a stable 2D manifold of a saddle point and the subsequent convergence towards a small region of the phase space (that the authors called a ``narrow door'') that triggers the extreme intensity pulses. Here, Figs.~\ref{fig:4}(a) and ~\ref{fig:4b}(a) show that there is a similar convergence of the trajectories towards a ``narrow door'' before extreme intensity pulses occur.

Figure~\ref{fig:4}(a) was done by performing one long simulation of the noise-less rate equations, selecting the pulses that are above a certain amplitude threshold (in this case equal to 10) and superposing sections of the intensity time-trace that contain the extreme pulse. Each section covers a time interval of 60 ns and the superposition is done by centering each section at the peak of the extreme pulse. One can observe that the superposition generates a narrow curve well before and after the extreme pulse occurs. This narrow curve is also observed in Fig. \ref{fig:4b}(a) that displays the superposition of sections of the trajectory (before and after the extreme pulse occurs) in the two-dimensional plane (phase delay, carrier density).  On the contrary, in Figs. ~\ref{fig:4}(b) and ~\ref{fig:4b}(b), which are done in the same way as Figs. \ref{fig:4}(a) and \ref{fig:4b}(a) but with a lower threshold (now equal to 6), the superposition of sections of the intensity time-trace containing a pulse above the 
threshold [Fig. ~\ref{fig:4}(b)] and the superposition of the corresponding sections of the trajectory [Fig. ~\ref{fig:4b}(b)] is considerable more disperse and does not generate a narrow curve.

These plots support a qualitative comparison of the behavior with feedback and with injection and, despite the differences in both systems, suggests that extreme intensity pulses in the optical feedback case could occur through a similar mechanism as in the optical injection case. However, the feedback time delay renders the phase space infinite dimensional and thus, the topological analysis of the trajectories in phase space is difficult.

\begin{figure}[]
\includegraphics[width=6cm]{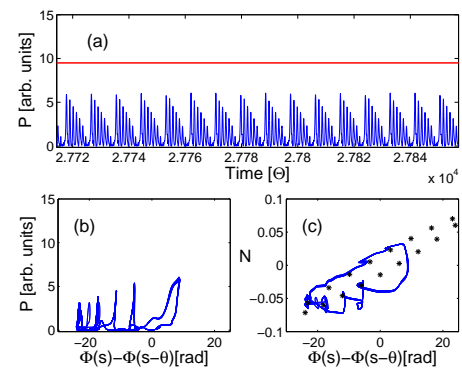}
\caption{Time series of the laser intensity just before {\it transition 2} ($\eta=0.073$), displaying regular pulse packages (RPPs). The horizontal line indicates the threshold for extreme pulses.\label{fig:RPPs}}
\end{figure}

With strong feedback the attractor suddenly disappears (at $\eta\sim 0.073$, see Fig.~\ref{fig:1} (a)) and the intensity becomes oscillatory with constant small amplitude. We will refer to this second abrupt change as {\it transition 2}. At this transition no sign of intermittency was observed. As shown in Fig.~\ref{fig:RPPs}, before the attractor is destroyed the dynamics is periodic and corresponds to regular pulse packages, which are well below the threshold of extreme pulses.

The new attractor is now followed by decreasing the feedback strength, and its bifurcations are shown in Fig.~\ref{fig:1} (b). The attractor described previously (referred to as attractor 1) coexists with this new attractor (referred to as attractor 2) in a wide range of feedback strengths, as could be expected for a dissipative system. The periodic intensity undergoes a Hopf bifurcation after which the cw regime is reached. Then, the trajectory remains in this steady state (corresponding to a stable ECM) until the ECM disappears after a saddle-node bifurcation occurring at $\eta \sim 0.066$. At this point, in the following referred to as {\it transition 3}, the trajectory evolves back to attractor 1, that sustains occasional extreme pulses, thus closing a hysteresis cycle. It seems that attractor 1 is not affected by the appearance/dissapearance (for increasing/decreasing $\eta$) of the ECM. This can be due the fact that attractor 1 and the ECM are well separated in the phase space, because attractor 1
originates from
a different ECM.

Once the trajectory is back in attractor 1, if we continue decreasing the feedback strength the bifurcation scenario is the same as that observed when we increased the feedback [compare Fig.~\ref{fig:1} (b) after {\it transition 3} with Fig.~\ref{fig:1} (a)], i. e. no hysteresis is observed at {\it transition 1}.

\begin{figure}[]
\includegraphics[width=6cm]{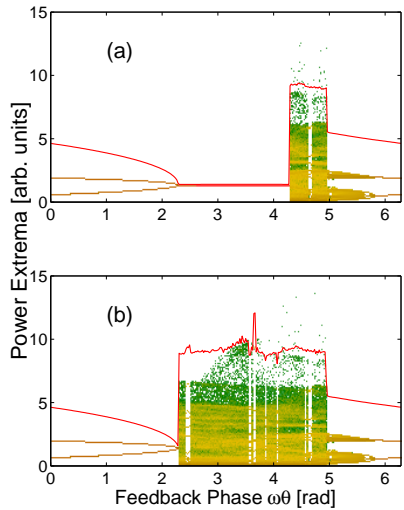}
\caption{Bifurcation diagram when increasing (a) and decreasing (b) the feedback phase. $\eta=0.0642$, other parameters as is Fig.~\ref{fig:1}. \label{fig:feedback_phase}}
\end{figure}

The bifurcation scenario described for varying feedback strength is also observed when varying the feedback phase. In Fig.~\ref{fig:feedback_phase} we observe three similar transitions when increasing and decreasing $w\theta$.

\subsection{Influence of spontaneous emission noise}

 \begin{figure*}[]
 \includegraphics[width=5.8 cm]{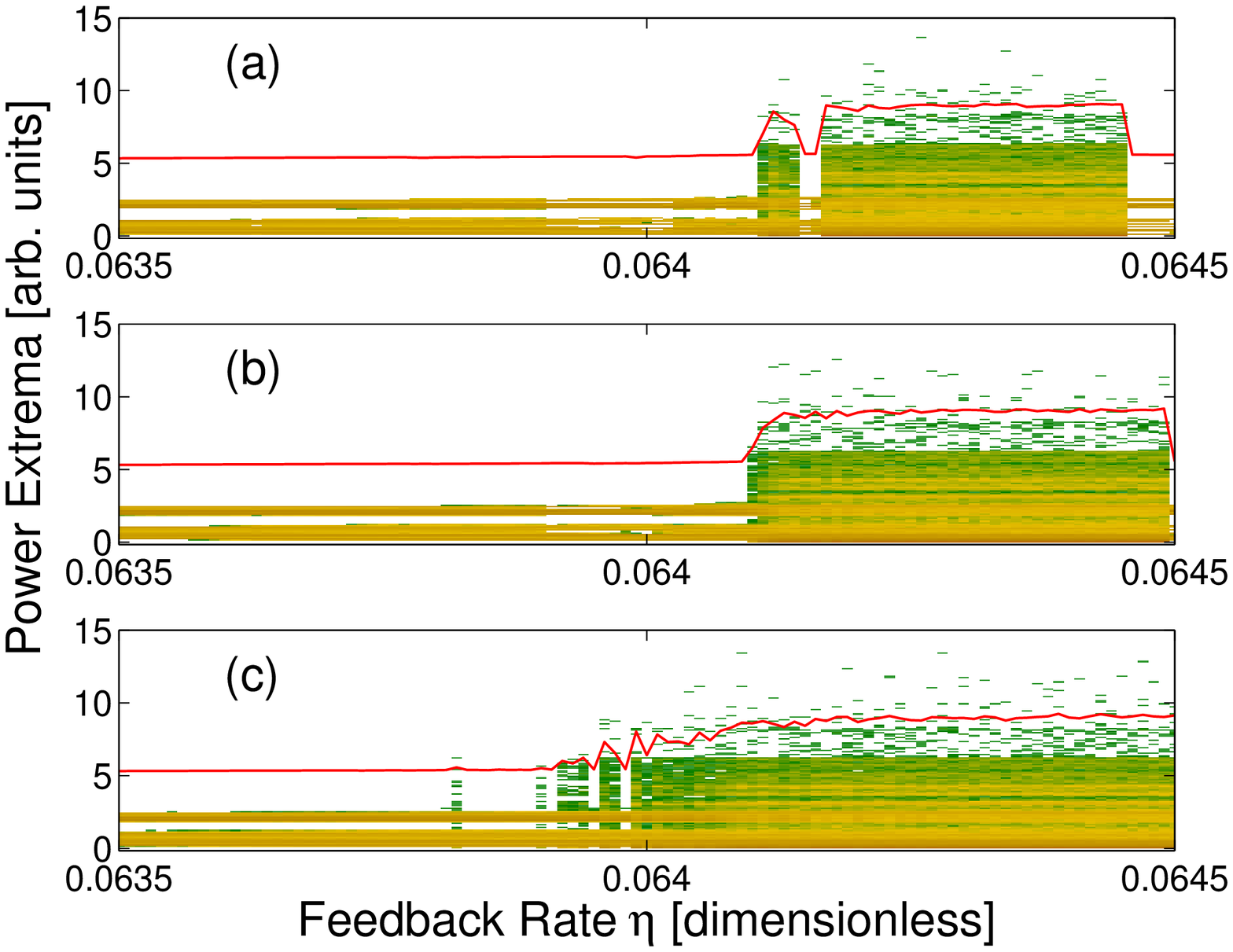}
 \includegraphics[width=5.8 cm]{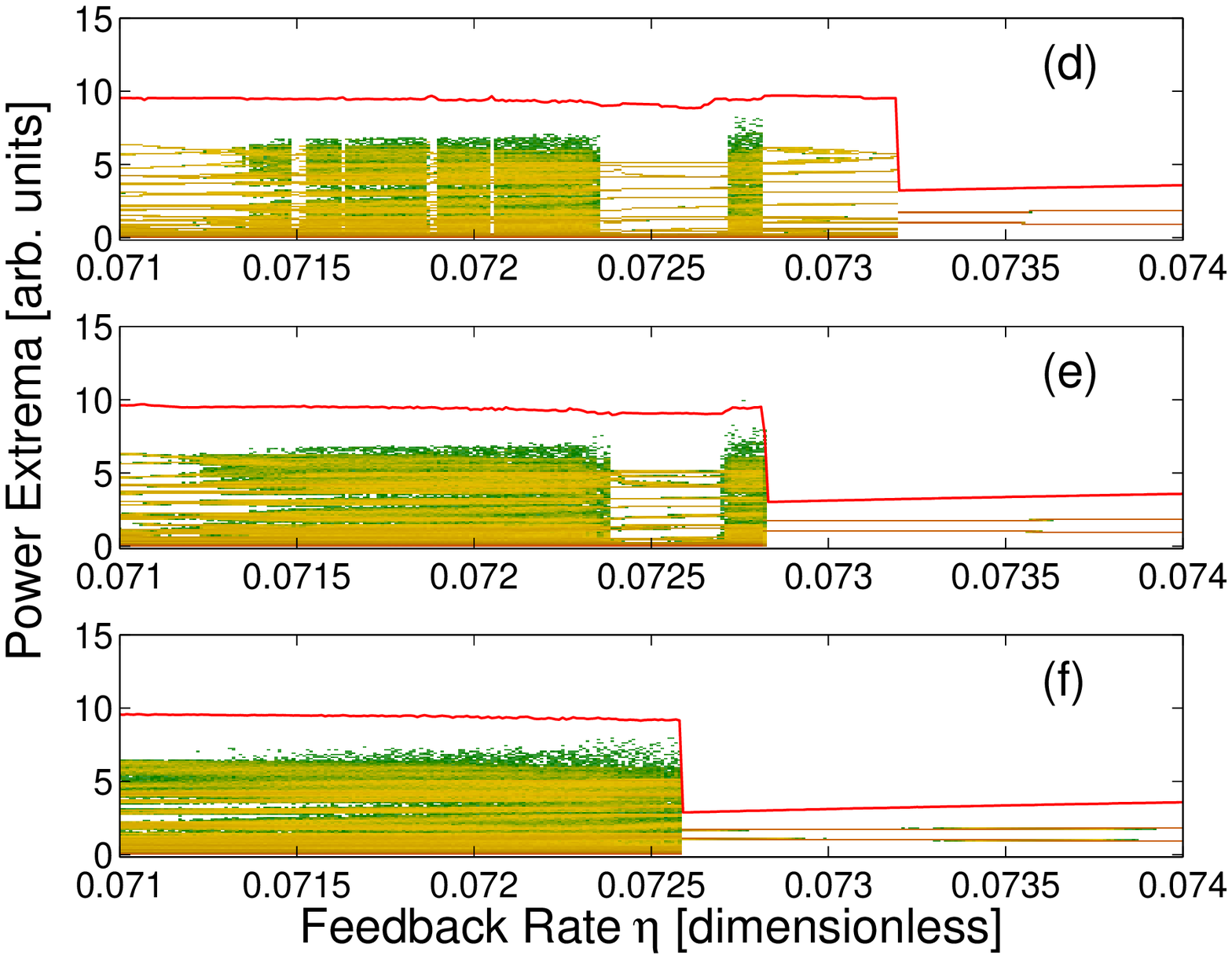}
 \includegraphics[width=5.8 cm]{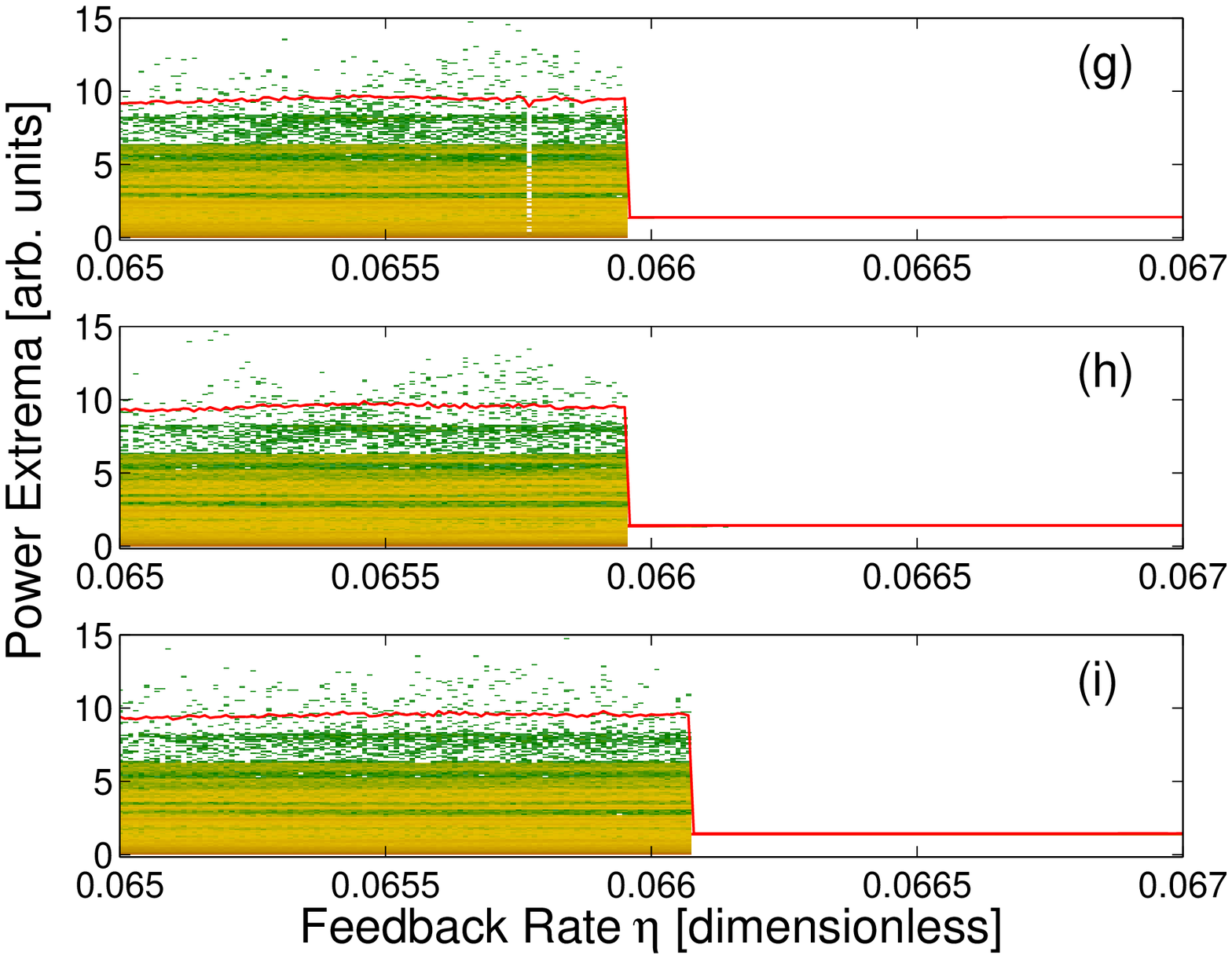}
 \caption{From (a) to (c) zoom of {\it transition 1} when the feedback rate increases. (d)-(f) zoom of {\it transition 2} when the feedback rate increases. (g)-(i) zoom of {\it transition 3} when the feedback rate decreases. The noise strength is $\beta=0$ (a,d,g), $10^{-4}$ (b,e,h) and $10^{-3}$ (c,f,i).}
 \label{fig:noise}
 \end{figure*}

One could expect that random fluctuations, unavoidable in lasers, could induce switchings between the coe\-xisting attractors previously discussed. In order to explore the effect of noise we did the bifurcation diagrams shown in Fig.~\ref{fig:1} sweeping the feedback strength upwards and downwards, but now including additive noise in the simulations. In Fig.~\ref{fig:noise} we show in detail the three transitions with and without noise.
Before {\it transition 1} noise leads to occasional extreme intensity pulses (Fig.~\ref{fig:noise} (a)-(c)). These noise-induced pulses are due to large excursions in the phase space, after which the trajectory relaxes back to the attractor.

The effect of noise observed at {\it transitions 2} and {\it 3} is the one expected when two attractors coexist: in Fig.~\ref{fig:noise} (d)-(f) (the feedback increases) one can notice that noise anticipates the transition to attractor 2, because {\it transition 2} occurs at a smaller value of $\eta$. In Fig.~\ref{fig:noise} (g)-(i), we see that {\it transition 3} (the feedback decreases) is also anticipated, i.e., occurs at a higher $\eta$ value.

\section{Conclusion and discussion}

To summarize, we studied numerically the dynamics of a semiconductor laser with optical feedback from a short external cavity and demonstrated the existence of para\-meter regions where the laser intensity displays extreme pulses. We identified three relevant transitions involved in the appearance of the extreme pulses. {\it Transition 1} and {\it transition 2} are related to the appearance and destruction of the attractor that sustains extreme pulses. This attractor coexist with another smaller one in a broad range of feedback strengths (from {\it transition 2} to {\it transition 3}) which defines a hysteresis cycle. We also demonstrated that noise does not modify this scenario but only anticipates the different transitions.

While our results are fully consistent with the findings of Ref.~\cite{pre_2004}, where the sequence of bifurcations leading to regular pulses packages (RPPs) was studied in detail, our simulations suggest that RPPs and extreme pulses are different dynamical behaviors and occurs in different parameter ranges. This is because the RPPs have a well defined periodicity and pulse amplitudes that are not extreme [Fig.~\ref{fig:RPPs}], and also, because even when strong noise is included in the simulations, RPPs are robust and no extreme pulses are observed for parameters corresponding to RPP dynamics [Fig.~\ref{fig:noise}(f)]. On the contrary, strong noise is capable of inducing extreme pulses for parameters before {\it transition 1} (at {\it transition 1} extreme pulses appear in the noiseless simulations) [Fig.~\ref{fig:noise}(c)].

Since diode lasers with integrated short external cavities are nowadays widely used in many applications that require single-frequency, compact and efficient light sources, our results can help avoiding extreme intensity pulses in these devices, that might originate due to un- controlled small variations of the feedback parameters.

{\it Acknowledgment.} This work was supported in part by grant FA8655-12-1-2140 from EOARD US, grant FIS2009-13360 from the Spanish MCI, and grant 2009 SGR 1168 from the Generalitat de Catalunya. C. Masoller acknowledges support from the ICREA Academia programme. J. Z. M. acknowledges support from the grant FISICOS of the Spanish MCI (FIS2007-60327). J. A. R. acknowledges support from grant BES-2008-003398 and thanks UPC hospitality during his visit, during which part of this work was done.


\begin{thebibliography}{99}
\bibitem{prl_2005} J. Masoliver, M. Montero and J. Perell\'o, Phys. Rev. E {\bf 71}, 056130 (2005).
\bibitem{pre_2012} C. Nicolis and G. Nicolis, Phys. Rev. E {\bf 85}, 056217 (2012).
\bibitem{pre_2011} T. Schweigler and J. Davidsen, Phys. Rev. E {\bf 84}, 016202 (2011).
\bibitem{cc_2010} Richard W. Katz, Climatic Change {\bf 100}, 71 (2010).
\bibitem{epj_2012} V.I. Yukalov, E.P. Yukalova and D. Sornette, Eur. Phys. J-Special Topics {\bf 205}, 313 (2012).
\bibitem{pre_2009} H. E. Roman, R. A. Siliprandi, C. Dose, and M. Porto, Phys. Rev. E {\bf 80}, 036114  (2009).
\bibitem{ol_2011} M.G. Kovalsky, A.A. Hnilo and J.R. Tredicce, Opt. Lett. {\bf 36}, 4449 (2011).
\bibitem{prl_2011} C. Bonatto, M. Feyereisen, S. Barland, M. Giudici, C. Masoller, Jose R. Rios Leite and J. R. Tredicce,
Phys. Rev. Lett. {\bf 107}, 053901 (2011).
\bibitem{el_2012} K. Schires, A. Hurtado, I.D. Henning and M.J. Adams, 
    Electronics Lett. {\bf 48}, 14 (2012).
\bibitem{ol_2013} A.K. Dal Bosco, D. Wolfersberger and M. Sciamanna, to appear in Opt. Lett. (2013).
\bibitem{prl_2001} T. Heil, I. Fischer, W. Els\"a{\ss}er and A. Gavrielides,
Phys. Rev. Lett. {\bf 87}, 243901 (2001).
\bibitem{pre_2004} A. Tabaka, K. Panajotov, I. Veretennicoff and M. Sciamanna, 
    Phys. Rev E  {\bf 70}, 036211 (2004).
\bibitem{pra_2006} A. Tabaka, M. Peil, M. Sciamanna, I. Fischer, W. Els\"a{\ss}er, H. Thienpont, I. Veretennicoff and K. Panajotov,
Phys. Rev. A {\bf 73}, 013810 (2006).
\bibitem{lk} R. Lang and K. Kobayashi, IEEE J. Quantum Electron. {\bf 16}, 347 (1980).
\bibitem{libro_deb_alan} D. M. Kane and K. A. Shore, {\it Unlocking dynamical diversity: optical feedback effects on semiconductor lasers} (Wiley, 2005).
\bibitem{libro_othsubo} J. Ohtsubo, {\it Semiconductor Lasers: Stability, Instability and Chaos, 2nd ed.} (Springer, Berlin, 2010).
\bibitem{RogueBook} C. Kharif, E. Pelinovsky, and A. Slunyaev, {\it Rogue Waves in the Ocean} (Springer, Heidelberg, 2009).
\bibitem{ExtremeBook} S. Albeverio, V. Jentsch, and H. Kantz, {\it Extreme Events in Nature and Society} (Center for Frontier Sciences, Germany, 2006).
\bibitem{sano} T. Sano,
Phys. Rev. A {\bf 50}, 2719 (1994).
\bibitem{masoller} C. Masoller and N. B. Abraham,
 Phys. Rev. A {\bf 57}, 1313 (1998).
\bibitem{gavrielides} R. L. Davidchack, Y-C Lai, A. Gavrielides and V. Kovanis
Physica D {\bf 145}, 130 (2000).
\bibitem{torcini} A. Torcini, S. Barland, G. Giacomelli, and F. Marin,
Phys. Rev. A {\bf 74}, 063801 (2006).
\bibitem{zamora} J. Zamora-Munt, C. Masoller and J. Garcia-Ojalvo,
Phys. Rev. A {\bf 81}, 033820 (2010).
\bibitem{epj_2010} K. Al-Naimee, F. Marino, M. Ciszak, S.F. Abdalah, R. Meucci and F.T. Arecchi, Eur Phys. J. D {\bf 58}, 187 (2010).
\bibitem{pra_2013} J. Zamora-Munt, B. Garbin, S. Barland, M. Giudici, J. R. Rios Leite, C. Masoller and J. R. Tredicce, Phys. Rev. A. {\bf 87}, 035802 (2013).

\end{thebibliography}
\end{document}